\documentstyle[aps,prl,multicol]{revtex}

\draft

\begin{document}

\title{Bloch electrons in a Jahn-Teller crystal \\
  and an orbital-density-wave state due to the Berry phase}

\author{Hiroyasu Koizumi}
\address{Faculty of Science, Himeji Institute of Technology,
  Kanaji, Kamigori, Ako-gun, Hyogo 678-12, Japan}

\author{Takashi Hotta and Yasutami Takada}
\address{Institute for Solid State Physics, University of Tokyo,
  7-22-1 Roppongi, Minato-ku, Tokyo 106, Japan}

\date{\today}

\maketitle
\begin{abstract}
The effect of the Berry phase is included explicitly in the
wavefunction describing conduction electrons in a crystal composed of
periodically arrayed Jahn-Teller centers that have conically
intersecting potential energy surfaces.
The Berry phase can make a drastic change in the band structure,
leading generally to the formation of an orbital-density-wave state.
We discuss implications of our theory and possible relations to the
orbital ordering observed in the manganese perovskites.
\end{abstract}


\begin{multicols}{2}
\narrowtext

The interplay among charge, spin, and orbital degrees of freedom is a
central issue in recent study on the perovskite manganites such as
La$_{1-x}$Ca$_x$MnO$_3$ \cite{ref1}.
In particular, the direct observation of the orbital ordering
\cite{ref2} makes the importance of the orbital degrees of freedom
even clearer.

In this work, we investigate Bloch electrons in a Jahn-Teller (JT)
crystal by focusing on the effects of double degeneracy of $e_g$
orbitals at each MnO$_6$ octahedron, and its lift due to the
Jahn-Teller (JT) distortions.
Two groups have already reported the effects of the JT distortions
on the electronic properties by implementing first-principles
band-structure calculations \cite{ref3a,ref3b}.
Although these calculations are the state of the art, both groups
overlook a very important ingredient of the problem, namely, the Berry
phase or the geometric phase \cite{ref4} associated with lattice
distortions in each JT center in a crystal.

In view of this situation, we have done the following: 
First, by starting with a rather general argument on a periodic array
of JT centers, we specify a simple model representing a linear chain
of MnO$_6$ octahedra.
On the basis of this model, we describe how the Bloch's theorem is
modified in the presence of the Berry phase.
Second, we show that its presence brings about a drastic change 
in the band structure compared to the result in its absence. 
This calls for revision of conventional band-structure 
calculations for materials involving the JT centers. 
Third, we find that the optimum JT distortions can be obtained 
by making the total energy minimum and that the resultant 
ground state exhibits an oscillation of the orbital density,
the orbital density wave (ODW). 
In particular, the ODW is reduced to the orbital-ordering state 
with an appropriate choice of parameters involved in the system. 
Fourth, we envisage that many experimental results 
for the perovskite manganites related to the orbital degrees 
of freedom may be explained in the light of our theory.

Consider a periodic array of JT centers. 
Here the word ``JT center'' is used in a slightly broader sense than
usual.
We use it for a complex of atoms whose vibronic states arise from the
interaction of degenerate electronic states and linearly-coupled
vibrational modes.
(Note that the degeneracy does not necessarily come from the
point-group symmetry.)
Such a complex has structural instability which causes distortions of
the atomic configuration, leading to the lift in the degeneracy of the
electronic states at equilibrium.
If more than two linearly-coupled vibrational modes are present,
potential energy surfaces exhibit so-called
``a conical intersection (CI)'' \cite{Yarkony96}.
The CI has also drawn much attention recently, due to the topological
phase arising from it. 
This is an example of the Berry phase, which has been known in the
context of dynamical JT effects for years \cite{LHH}.

We specify the JT center we consider in this paper.
It is composed of two reference electronic states, $\psi_a$ and
$\psi_b$, and two vibrational modes, $Q_a$ and $Q_b$, that couple with 
the electronic states linearly. 
In matrix representation using $\{\psi_a, \psi_b \}$ as a basis, the
vibronic interaction is given by 
\begin{eqnarray}
  \left(
    \begin{array}{cc}
      \alpha Q_a & \beta Q_b \\
      \beta Q_b & \gamma Q_a
    \end{array}
  \right).
  \label{eq:VI}
\end{eqnarray}
This matrix is diagonalized to yield eigenvalues, 
$E^{\pm} = \{(\alpha + \gamma)Q_a 
\pm [(\alpha-\gamma)^2 Q_a^2 +4\beta^2 Q_b^2]^{1/2}\}/2$, 
which possess a conically intersecting degeneracy point at
$Q_a=Q_b=0$, if $\alpha \neq \gamma $.

The isolated JT center described above is modified in a crystal.
However, the CI seen in $E^{\pm}$ survives against a perturbation
preserving the time-reversal invariance \cite{Longuet-Higgins75}:
Application of such a perturbation modifies the vibronic coupling
matrix into
\begin{eqnarray}
  \left(
    \begin{array}{cc}
      \alpha Q_a +a & \beta Q_b +b \\
      \beta Q_b +b  & \gamma Q_a+c 
    \end{array}
  \right),
\end{eqnarray}
where real numbers $a$, $b$, and $c$ represent perturbations from
crystal environments, but this merely shifts the CI position to
$Q_a=(c-a)/(\alpha-\gamma), Q_b=-b/\beta$ \cite{comment}.

Now we construct a model Hamiltonian $H$ for the crystal composed of
the JT centers. 
Using the vibronic interaction expressed in Eq.~(\ref{eq:VI}), $H$ is
given by
\begin{eqnarray}
  \label{H}
  H&=&-\sum_{\langle i,j \rangle}
  (t_{1}a_i^{\dag}a_j+t_{2}b_i^{\dag}b_j
  + t_{3}a_i^{\dag}b_j + t_{3}b_i^{\dag}a_j + {\rm h.c.}) 
  \nonumber \\
  &&+ \sum_j \bigl [
  (\alpha Q_{aj}+a) a_j^{\dag}a_j 
  +(\gamma Q_{aj}+c) b_j^{\dag}b_j 
  \nonumber \\
  &&~ \qquad +(\beta Q_{bj}+b)(a_j^{\dag}b_j+b_j^{\dag}a_j)\bigr ],
\end{eqnarray}
where the first sum describes the electron transfer effects between
the JT centers and $\langle i,j \rangle$ indicates a nearest-neighbor
pair.
The second sum is the electronic Hamiltonian part for the JT center at
site $j$, where $Q_{aj}$ and $Q_{bj}$ are $Q_a$ and $Q_b$ at site $j$,
respectively, and $a_j$ and $b_j$ are, respectively, annihilation
operators for electrons in the states $\psi_a$ and $\psi_b$.
In the later discussion on manganese oxides, $a_j$ and $b_j$ are,
respectively, identified as the electronic states steming from
$d_{x^2-y^2}$ and $d_{3z^2-r^2}$ orbitals of Mn$^{3+}$, which are
hybridized with oxygen orbitals.

In Eq.~(\ref{H}), the spin degree of freedom is neglected for
simplicity, but this is legitimate for the perovskite manganites due
to the Hund's coupling to the electrons in the $t_{2g}$ orbitals.
As for lattice vibrations, we have not included the unperturbed phonon
Hamiltonian $H_{ph}$ which treats the restoring forces for $Q_{aj}$
and $Q_{bj}$ as well as the kinetic-energy parts.
In this paper, we restrict ourselves to the situation in which the
electron density per site is constant at all JT centers.
Then $H_{ph}$ is reduced to a constant and thus it can be neglected.

Let us transform the electronic basis from $\{a_j, b_j \}$ into 
$\{c_j, d_j \}$, where $c_j \equiv (a_j+ib_j)/\sqrt{2}$ and 
$d_j \equiv (a_j-ib_j)/\sqrt{2}$. 
In this new basis, $H$ is rewritten as 
\begin{eqnarray}
  H&=&-\sum_{\langle i,j \rangle}(tc_i^{\dag}c_j+td_i^{\dag}d_j
  + s c_i^{\dag}d_j + s^{\ast}d_i^{\dag}c_j + {\rm h.c.}) \nonumber \\
  &&+ \sum_j [U_j(c_j^{\dag}c_j+d_j^{\dag}d_j)
  +V_j c_j^{\dag}d_j+ V_j^{\ast}d_j^{\dag}c_j ],
  \label{eq:Hamil}
\end{eqnarray}
where $t=(t_1+t_2)/2$, $s=(t_1-t_2)/2+it_3$,
$U_j = [(\alpha + \gamma )Q_{aj}+a+c]/2$, and 
$V_j = [(\alpha-\gamma)Q_{aj}+a-c]/2 -i(\beta Q_{bj}+b)$.
Note that the circular motion of $Q_{aj}$ and $Q_{bj}$ around the CI
point $(Q_{aj},Q_{bj})=((c-a)/(\alpha-\gamma),-b/\beta)$ changes the
phase of $V_j$.
This phase change is what responsible for the appearance of the Berry
phase. 
Because we assume that the electron density per site is constant at
all JT centers, the absolute value of $V_j$ is independent of $j$,
though its phase can vary from site to site.

For the $e_g$ electrons in the MnO$_6$ octahedron, the sum of $\alpha$
and $\gamma$ vanishes by symmetry. 
This makes the parameters $U_j$ constant and these parameters can be
removed by shifting the origin of energies.
For this reason, we will not be concerned with the $U_j$ terms.

We discuss the modification of the Bloch's theorem based on the model
in Eq.~(\ref{eq:Hamil}) in one spatial dimension with $t_1=t_2$ and
$t_3=0$ for the time being.
This simplification leads us to an analytic expression without loss of
essential physics.
In this choice of parameters, $H$ in momentum representation is
reduced to
\begin{eqnarray}
  H &=& \sum_{k} \varepsilon_k
  (c_{k}^{\dagger}c_{k}+d_{k}^{\dagger}d_{k}) \nonumber \\ 
  &&+ \sum_{k,q} (V_q c_{k+q}^{\dagger}d_{k}
  + V_q^* d_{k}^{\dagger}c_{k+q}),
\end{eqnarray}
where $\varepsilon_k= -2t\cos k$ and 
$V_q = (|V|/N) \sum_{j}e^{-i(q j-\xi_j)}$.
Here we set the lattice constant as unity and $N$ is the total number
of sites.

Although it appears in $V_q$, the phase $\xi_j$ does not play a role,
if it is independent of site, i.e., $\xi_j=\xi$.
In fact, the one-electron eigen energy for $H$ is given by 
$E_k^{\pm} = \varepsilon_k \pm |V|$ in this case and the corresponding
eigenfunction $\varphi_k^{\pm}(x)$ obeys the Bloch's theorem,
characterized by the property
\begin{equation}
  T \varphi^{\pm}_k(x)=\varphi^{\pm}_k(x+1)=e^{ik}\varphi^{\pm}_k(x),
\end{equation}
with $T$ the operator to translate the whole system by a unit lattice
vector.

Imagine now that $\xi_j$ changes from $0$ to $2\pi$ site by site in
the period of $M$ JT centers. 
For simplicity, we assume that $\xi_j = 2\pi j /M$ and that $M$ is
commensurate with $N$.
In this situation, $V_q$ is given by $V_q= |V|\delta_{q,2\pi/M}$ and
the one-electron eigen energy is determined as
\begin{eqnarray}
  E_{k}^{\pm} =
  [\varepsilon_{k}+{\tilde \varepsilon}_{k} \pm
  \sqrt{ (\varepsilon_{k}-{\tilde \varepsilon}_{k})^2+4|V|^2}]/2, 
\end{eqnarray}
with ${\tilde \varepsilon}_k=\varepsilon_{k+2\pi/M}$.
The corresponding eigenfunction $\varphi_k^{\pm}(x)$ satisfies the
modified Bloch's theorem:
\begin{equation}
  \label{Bloch}
  T 
  \left[
    \begin{array}{l}
      \varphi_k^{+}(x) \\ \varphi_k^{-}(x)
    \end{array}
  \right] = 
  e^{i(k+\pi/M)}UPU^{-1}
  \left[
    \begin{array}{l}
      \varphi_k^{+}(x) \\ \varphi_k^{-}(x)
    \end{array}
  \right] ,
\end{equation}
where the matrices $P$ and $U$ are defined as
\begin{equation}
  P =
  \left(
    \begin{array}{cc}
      e^{i\pi/M} & 0 \\
      0 & e^{-i\pi/M}
    \end{array}
  \right), \quad
  U=
  \left( 
    \begin{array}{cc}
      p_k^+ & p_k^- \\ 
      -p_k^- & p_k^+ 
    \end{array}
  \right),
\end{equation}
with $p_k^{\pm}$, given by 
\begin{eqnarray}
  p_{k}^{\pm} = 
  \Biggl[ {1 \over 2} \pm 
  \frac{{\tilde \varepsilon}_{k}-\varepsilon_{k}}
  {2\sqrt{({\tilde \varepsilon}_{k}-\varepsilon_{k})^2 + 4|V|^2}}
  \Biggr]^{1/2}.
\end{eqnarray}
In addition to the ordinary phase factor $e^{ik}$, the Bloch function
acquires the Berry phase.
It takes account of the difference in the electron motion around the
JT center, clockwise or counterclockwise, indicating that careful
treatments are needed for the analysis of a system including JT
centers.

The phase factor $e^{i\pi/M}$ in Eq.~(\ref{Bloch}) assures
single-valuedness of basis functions which satisfy 
$T^M  \varphi^{\pm}_k(x) = e^{iMk}\varphi^{\pm}_k(x)$ 
in accordance with the Bloch's theorem in a crystal with the lattice
constant $M$; without it, signs of basis functions change after the
translation by an odd $M$ number of sites.
This factor may be considered as one arising from 
``a ficticious magnetic field'' that exists at each CI \cite{Mead80}.
On the other hand, $P$ accounts for the site-to-site variation of the
lattice-twisting phase $\xi_j$.

The dispersion curves representing $E_{k}^{\pm}$ are plotted in an
extended Brillouin zone for various values of $M$ in Fig.~\ref{fig1}
in which $|V|/t$ is taken to be 0.5 and the origin of momenta is
shifted by $-\pi/M$ in each panel.
This shift is equivalent to use $K=k+\pi/M$ instead of $k$, where $K$
is the generalized quasi-momentum and $\pi/M$ arises from the
ficticious magnetic field \cite{Lifshitz80}.

For $M=\infty$ (or $M=1$), the dispersion relation 
$E_k^{\pm} = \varepsilon_k \pm |V|$ is the one corresponding to the
result in the conventional band-structure calculation
\cite{ref3a,ref3b}.
For $M=2$, we observe the same band gap with the magnitude of $2|V|$
as that for $M=\infty$, but the dispersion curve itself is completely
different. 
We can even predict an insulating state at ``half-filling'', where the
electron density per site $n$ is unity.
For the case of $2 < M < \infty$, we also find a drastic change in
$E_{k}^{\pm}$ from that for $M=\infty$, but the system remains
metallic irrespective of $n$.

For given $n$, $M$ and $|V|/t$, information on the dispersion relation
allows us to evaluate the ground-state energy $E_M(V)$ by filling
electrons in the one-electron states from the bottom.
The relative stability among the states with different $M$'s can be
determined by comparing these $E_M(V)$'s.
A typical result is shown for $|V|/t=0.5$ in Fig.~\ref{fig2} in which
$\delta E(V,M)$, defined by $\delta E(V,M) \equiv E_M(V)-E(V=0)$, are
given as a function of $n$.
Note that at $V=0$, the ground-state energy is independent of $M$.
Thus $E(V=0)$ is written without the suffix $M$.

The result in Fig.~\ref{fig2} indicates that each $M$ has a
characteristic electron filling, $n(M)$, at which $\delta E(V,M)$
takes a local minimum.
For an arbitrary $V$, $n(M)$ can be evaluated only numerically, but
for $|V|/t \ll 1$ and small $M$, it turns out that $n(M)$ is well
approximated by $n(M)=2/M$ (and $n(M)=2-2/M$ due to the particle-hole
symmetry).
The comparison of $\delta E(V,M)$'s suggests that the JT-distorted
state with the period $M$ is expected to occur spontaneously for $n$
around $n(M)$.
Note that $\partial E/\partial n$ exhibits a jump at $n=n(M)$ only for
$M=2$ due to the insulating nature.
For other $M$'s, no jump can be seen at $n=n(M)$.

 From the viewpoint of our theory, we now try to understand the orbital
ordering observed in La$_{1-x}$Ca$_x$MnO$_3$.
We concentrate our interest into the case of $x=0$ or LaMnO$_3$ which
corresponds to the half-filling in our model.
At $n=1$, the above discussion on the ground-state energy implies that
a periodic JT-distorted state with $M=2$, or an ODW state composed of
bipartite $A$ and $B$ sublattices, is stabilized.
This agrees qualitatively well with the experimental situation.

As for the orbital pattern, LaMnO$_3$ exhibits an orbital order
between the orbitals of $d_{3x^2-r^2}$ and $d_{3y^2-r^2}$ rather than
$d_{x^2-y^2}$ and $d_{3z^2-r^2}$.
In order to see whether this pattern can be predicted in our theory or
not, we take a more realistic choice of parameters.
For this purpose, we set the hopping energies as 
$t_1=3t_0/4$, $t_2=t_0/4$, and $t_3=\sqrt{3}t_0/4$ by assuming 
that both $\psi_a$ and $\psi_b$ are purely $d$-wave-like.
Here the energy unit $t_0$ is the overlap integral between adjacent
$d_{3z^2-r^2}$ orbitals situated in the $z$-direction.
In this choice, $t$ and $s$ are, respectively, given by $t=t_0/2$ and
$s=(t_0/2){\rm exp}(i\pi/3)$.
On the other hand, we set $|V|$ as $t_0$ by referring to the result of
band-structure calculations \cite{ref3a} in which the band gap due to
the JT distortions, $2|V|$, is seen to be about the same as the
bandwidth of the $e_g$ electrons.

With these parameters, we can determine $\xi_j$ at $A$ and $B$
sublattices, $\xi_A$ and $\xi_B$, by optimizing the ground-state
energy, leading to $\xi_A=\pi/3$ and $\xi_B=4\pi/3$
(or $\xi_A=4 \pi/3$ and $\xi_B=\pi/3$).
The phase difference, $|\xi_A-\xi_B|=\pi$ reflects the Berry-phase 
effect which is fixed by the condition $M=2$, while the phase $\pi/3$
takes account of the local phase due to the complex number $s$.

In order to see the orbital ordering in real space described by the
ground state obtained in this way, we evaluate the following
quantities:
\begin{eqnarray}
  \rho_{3x^2-r^2}(j)
  = \langle \phi_{xj}^{\dagger} \phi_{xj} \rangle, \quad 
  \rho_{3y^2-r^2}(j) 
  = \langle \phi_{yj}^{\dagger} \phi_{yj} \rangle,
\end{eqnarray}
where $\phi_{xj}$ and $\phi_{yj}$ are, respectively, the annihilation
operators for electrons in $d_{3x^2-r^2}$ and $d_{3y^2-r^2}$ orbitals
at site $j$.
These operators are given by the linear combination of $a_j$ and $b_j$
through $\phi_{xj}=\cos (2\pi/3)a_j+\sin (2\pi/3)b_j$ and
$\phi_{yj}=\cos (4\pi/3)a_j+\sin (4\pi/3)b_j$.

In Fig.~\ref{fig3}, the calculated results for $\rho_{3x^2-r^2}$ and
$\rho_{3y^2-r^2}$ are shown as a function of the site index $j$.
The alternating pattern between $d_{3x^2-r^2}$ and $d_{3y^2-r^2}$
orbitals emerges.
Thus we may call the present ODW state as
``the orbital-ordering state''.
Note that ``the orbital-ordering state'' is obtained without making
any fine tuning of parameters.
In fact, the obtained ODW state is not easily destroyed by changing
the hopping parameters around the present values.
However, it is destroyed completely, if we do not pay due attention to
the Berry phase, as shown in the lower panel in which we take
$\xi_A=\xi_B=0$.

We have also obtained following speculations:
First, the interband transitions between $E_k^{-}$ and $E_k^{+}$
are radically different from $M=\infty$ to $M=2$.
In the former, no spectral band is expected but it is expected in the
latter.
This may be connected with the observed anomalous optical spectra 
\cite{Okimoto}. 
Second, because the Berry phase represents a fictitious magnetic
field, application of ``real'' magnetic fields provides a profound
effect on the ODW.
This may be the key to explain the observed successive phase
transitions in magnetic fields \cite{Kawano}.
These facts among others may support a Berry-phase scenario to
understand the orbital ordering in the manganese perovskites,
although we need to make our model more realistic to compare with the
real situations.

In conclusion, we have demonstrated the importance of the Berry phase
in the calculation of band structures for crystals containing JT
centers.

HK thanks S. Sugano for continuous encouragements on this work.
YT acknowledges the support from the Mitsubishi Foundation.


\begin{figure}
  \caption{Dispersion curves for several values of $M$ with
    $|V|/t=0.5$. We shift the origin of momenta by $-\pi/M$ in each
    panel.}
  \label{fig1}
\end{figure}

\begin{figure}
  \caption{Stabilization energy of the $M$-period JT distorted state
    as a function of the electron filling $n$ for $|V|/t=0.5$.
    We note that $\delta E_M(V)$ is in proportion to $|V|^2$ for
    $|V|/t \ll 1$ and the proportional coefficient depends on $M$.}
  \label{fig2}
\end{figure}

\begin{figure}
  \caption{Orbital densities, $\rho_{3x^2-r^2}$ and $\rho_{3y^2-r^2}$,
    as a function of site $j$ for $t=t_0/2$ and 
    $s=(t_0/2){\rm exp}(i\pi/3)$.
    In the upper panel, $\xi_A$ and $\xi_B$ are determined by
    optimizing the ground-state energy, while in the lower panel, 
    we set $\xi_A=\xi_B=0$.}
  \label{fig3}
\end{figure}

\end{multicols}

\begin{references}
\bibitem{comment} If the perturbation breaks the time-reversal
  invariance, the off-diagonal elements usually aquire imaginary
  parts, resulting in the disappearance of the CI.
\end{references}
\end{document}